\documentclass[12pt]{article}
\usepackage{amsthm,amsmath,latexsym,amssymb,amsfonts,amscd}
\usepackage{graphics,lscape,fancyhdr,array,stmaryrd,euscript,wrapfig}
\usepackage{ifthen,empheq,slashed}
\usepackage{ifpdf}
\usepackage{verbatim,slashed}
\numberwithin{equation}{section}
\usepackage{hyperref,setspace}

\usepackage{mathrsfs}
\usepackage[numbers,sort&compress]{natbib}
\usepackage[nottoc]{tocbibind}

\newcommand{\be}{\begin{equation}}
\newcommand{\ee}{\end{equation}}

\newcommand{\fud}[2]{{}^{#1}{}_{#2}\,}
\newcommand{\fdu}[2]{{}_{#1}{}^{#2}\,}

\newcommand{\besubeqs}{\begin{subequations}}
\newcommand{\esubeqs}{\end{subequations}}

\usepackage{tensor}

\renewcommand{\bar}[1]{\overline{#1}}

\makeatletter
\newcommand{\subalign}[1]{%
  \vcenter{%
    \Let@ \restore@math@cr \default@tag
    \baselineskip\fontdimen10 \scriptfont\tw@
    \advance\baselineskip\fontdimen12 \scriptfont\tw@
    \lineskip\thr@@\fontdimen8 \scriptfont\thr@@
    \lineskiplimit\lineskip
    \ialign{\hfil$\m@th\scriptstyle##$&$\m@th\scriptstyle{}##$\crcr
      #1\crcr
    }%
  }
}
\makeatother

\newcommand{\TikzRect}[2]{\filldraw[color=black,fill=red]  (#1-\R,#2-\R) rectangle (#1+\R,#2+\R);}
\newcommand{\TikzRectB}[2]{\filldraw[color=black,fill=blue]  (#1-\R,#2-\R) rectangle (#1+\R,#2+\R);}

\newcommand{\FIELDS}{
\begin{tikzpicture}[scale=0.7]
\tikzset{%
  >=latex, 
  inner sep=0pt,%
  outer sep=2pt,%
  mark coordinate/.style={inner sep=0pt,outer sep=0pt,minimum size=3pt,
    fill=black,circle}%
}
\def\R{0.15}
\def\mar{0.15}

\draw[->,black,thick] (0,0) -- (0,11) node[left]{$\# A'$ };
\draw[->,black,thick] (0,0) -- (15,0) node[right]{$\# A$} ;

\filldraw[color=black,fill=green]  (14,6) circle (\R);
\node[right] at (14.3,6) {one-forms, $\omega$};

\TikzRect{14}{7}  
\node[right] at (14.3,7) {zero-forms, $C$};

\filldraw[color=black,fill=green]  (3,0) circle (\R);
\filldraw[color=black,fill=green]  (2,1) circle (\R);
\filldraw[color=black,fill=green]  (1,2) circle (\R);
\filldraw[color=black,fill=green]  (0,3) circle (\R);

\filldraw[color=black,fill=green]  (3,1) circle (\R);
\filldraw[color=black,fill=green]  (2,2) circle (\R);
\filldraw[color=black,fill=green]  (1,3) circle (\R);

\filldraw[color=black,fill=green]  (4,1) circle (\R);
\filldraw[color=black,fill=green]  (3,2) circle (\R);
\filldraw[color=black,fill=green]  (2,3) circle (\R);
\filldraw[color=black,fill=green]  (1,4) circle (\R);

\filldraw[color=black,fill=green]  (4,2) circle (\R);
\filldraw[color=black,fill=green]  (3,3) circle (\R);
\filldraw[color=black,fill=green]  (2,4) circle (\R);

\filldraw[color=black,fill=green]  (5,2) circle (\R);
\filldraw[color=black,fill=green]  (4,3) circle (\R);
\filldraw[color=black,fill=green]  (3,4) circle (\R);
\filldraw[color=black,fill=green]  (2,5) circle (\R);

\draw[green,thick] (3,0) -- (0,3) -- (2,5) -- (5,2) -- (3,0);

    \draw[-latex,thick] (4.5,6) node[right,scale=1.0]{$\omega^{A(t-1),A'(2s-t-1)}$}
        to[out=-180,in=50] (2.15, 5.15);
        
    \draw[-latex,thick] (5,8) node[right,scale=1.0]{$C^{A(t-1),A'(2s-t+1)}$}
        to[out=-180,in=0] (2.15,7);

    \draw[-latex,thick] (9,2.5) node[right,scale=1.0]{$C^{A(2s-t+1),A'(t-1)}$}
        to[out=180,in=-30] (7.2, 2);

    \draw[-latex,thick] (4.5,4) node[right,scale=1.0]{$\omega^{A(2s-t-1),A'(t-1)}$}
        to[out=180,in=90] (5, 2.15);

    \draw[red,thick] (2,11) -- (0,9) -- (2,7) -- (5,10);
    \TikzRect{2}{7}

    \draw[red,thick] (11,2) -- (9,0) -- (7,2) -- (10,5);
    \TikzRect{7}{2}
    
    \draw[rounded corners] ( 2-0.5 , 5-0.5 ) rectangle (2.5,7.5) {}; 
    \draw[rounded corners] ( 5-0.5 , 2-0.5 ) rectangle (7.5,2.5) {}; 
    
    \draw (-0.1,9) -- (0.1,9);
    \node[left] at (-0.1,9) {\footnotesize$2s$};
    \draw[dashed] (-0.1,7) -- (2,7);
    \node[left] at (-0.1,7) {\footnotesize$2s-t+1$};
    \draw[dashed] (-0.1,5) -- (2,5);
    \node[left] at (-0.1,5) {\footnotesize$2s-t-1$};
    \node[left] at (-0.1,3) {\footnotesize$2s-2t$};
    
    \draw[dashed] (7,2) -- (7,-0.1);
    \node[below] at (7,-0.1) {\footnotesize$2s-t+1$};
    \draw (9,0.1) -- (9,-0.1);
    \node[below] at (9,-0.1) {\footnotesize$2s$};
    \draw[dashed] (5,2) -- (5,-0.7);
    \node[below] at (5,-0.7) {\footnotesize$2s-t-1$};
    \node[below] at (3,-0.1) {\footnotesize$2s-2t$};
\end{tikzpicture}}

\newcommand{\FIELDSDescr}{
\begin{tikzpicture}[scale=0.7]
\tikzset{%
  >=latex, 
  inner sep=0pt,%
  outer sep=2pt,%
  mark coordinate/.style={inner sep=0pt,outer sep=0pt,minimum size=3pt,
    fill=black,circle}%
}
\def\R{0.15}
\def\mar{0.15}

\draw[->,black,thick] (0,0) -- (0,11) node[left]{$\# A'$ };
\draw[->,black,thick] (0,0) -- (15,0) node[right]{$\# A$} ;

\filldraw[color=black,fill=green]  (14,6) circle (\R);
\node[right] at (14.3,6) {one-forms, $\omega$};

\TikzRect{14}{7}  
\node[right] at (14.3,7) {zero-forms, $C$};

\filldraw[color=black,fill=green]  (7,0) circle (\R);
\filldraw[color=black,fill=green]  (6,1) circle (\R);
\filldraw[color=black,fill=green]  (5,2) circle (\R);
\filldraw[color=black,fill=green]  (4,3) circle (\R);
\filldraw[color=black,fill=green]  (3,4) circle (\R);
\filldraw[color=black,fill=green]  (2,5) circle (\R);
\filldraw[color=black,fill=green]  (1,6) circle (\R);
\filldraw[color=black,fill=green]  (0,7) circle (\R);

\draw[green,thick] (7,0) -- (0,7);

    \draw[-latex,thick] (4.5,6) node[right,scale=1.0]{$\omega^{A(t-1),A'(2s-t-1)}$}
        to[out=-180,in=50] (2.15, 5.15);
        
    \draw[-latex,thick] (5,8) node[right,scale=1.0]{$C^{A(t-1),A'(2s-t+1)}$}
        to[out=-180,in=0] (2.15,7);

    \draw[-latex,thick] (9,2.5) node[right,scale=1.0]{$C^{A(2s-t+1),A'(t-1)}$}
        to[out=180,in=-30] (7.2, 2);

    \draw[-latex,thick] (5,-1) node[left,scale=1.0]{$\omega^{A(2s-t-1),A'(t-1)}$}
        to[out=0,in=-90] (5, 1.85);

    \draw[red,thick] (0,9) -- (9,0);
    \TikzRect{2}{7}
    \TikzRect{1}{8}
    \TikzRect{0}{9}
    \TikzRect{3}{6}
    \TikzRect{4}{5}
    \TikzRect{5}{4}
    \TikzRect{6}{3}
    \TikzRect{7}{2}
    \TikzRect{8}{1}
    \TikzRect{9}{0}

    \TikzRect{7}{2}
    
    \draw[rounded corners] ( 7-0.5 , -0.5 ) rectangle (9.5,.5) {};
     \draw[rounded corners] ( -0.5 , 7-0.5 ) rectangle (.5,9.5) {};    
    
    \draw[rounded corners] ( 2-0.5 , 5-0.5 ) rectangle (2.5,7.5) {}; 
    \draw[rounded corners] ( 5-0.5 , 2-0.5 ) rectangle (7.5,2.5) {}; 
\end{tikzpicture}}

\usepackage{tikz}
\usetikzlibrary{patterns}
\newcommand{\FIELDSBeyond}{
\begin{tikzpicture}[scale=0.7]
\tikzset{%
  >=latex, 
  inner sep=0pt,%
  outer sep=2pt,%
  mark coordinate/.style={inner sep=0pt,outer sep=0pt,minimum size=3pt,
    fill=black,circle}%
}
\def\R{0.15}
\def\mar{0.15}

\draw[->,black,thick] (0,0) -- (0,11) node[left]{$\# A'$ };
\draw[->,black,thick] (0,0) -- (15,0) node[right]{$\# A$} ;

    \draw[red,thick] (2,11) -- (0,9);
    \draw[red,dashed] (0,9) -- (6,3);
    \draw[red, thick] (6,3) -- (11,8);
    \TikzRect{6}{3}

    \draw[blue,thick] (11,2) -- (9,0);
    \draw[blue,dashed] (9,0) -- (3,6);
    \draw[blue,thick] (3,6) -- (8,11);
    \TikzRectB{3}{6}
    
    \draw[pattern=north west lines, pattern color=gray]
    (8,11) -- (3,6) -- (6,3) -- (11,8);
    
    \draw (-0.1,9) -- (0.1,9);
    \node[left] at (-0.1,9) {\footnotesize$2s$};
    
    \draw (9,0.1) -- (9,-0.1);
    \node[below] at (9,-0.1) {\footnotesize$2s$};
    
    \draw[dashed] (6,3) -- (6,-0.1);
    \node[below] at (6,-0.1) {\footnotesize$s+k+1$};
    \draw[dashed] (3,6) -- (3,-0.1);
    \node[below] at (3,-0.1) {\footnotesize$s-k+1$};
\end{tikzpicture}}

\RequirePackage{genyoungtabtikz}
\Yboxdim{6pt}
\Ylinethick{1pt}

\begin{document}
\pagenumbering{gobble}
\hfill
\vskip 0.01\textheight
\begin{center}
{\Large\bfseries 
Chiral approach to partially-massless fields}

\vskip 0.03\textheight
\renewcommand{\thefootnote}{\fnsymbol{footnote}}
Thomas \textsc{Basile}${}^a$, Shailesh \textsc{Dhasmana}${}^a$ \& Evgeny \textsc{Skvortsov}\footnote{Research Associate of the Fund for Scientific Research -- FNRS, Belgium}${}^{a}$\footnote{Also on leave from Lebedev Institute of Physics, Moscow, Russia}
\renewcommand{\thefootnote}{\arabic{footnote}}
\vskip 0.03\textheight

{\em ${}^{a}$ Service de Physique de l'Univers, Champs et Gravitation, \\ Universit\'e de Mons, 20 place du Parc, 7000 Mons, 
Belgium}\\

\begin{abstract}
    We propose a new (chiral) description of partially-massless
    fields in $4d$, including the partially-massless graviton,
    that is similar to the pure connection formulation
    for gravity and massless higher spin fields, the latter having
    a clear twistor origin. The new approach allows us to construct
    complete examples of higher spin gravities with (partially-)massless
    fields that feature Yang--Mills and current interactions. 
\end{abstract}

\end{center}
\newpage

\pagenumbering{arabic}
\setcounter{page}{1}

\section*{Introduction}
Partially-massless fields constitute a novel class of fields
that appears in the presence of a non-vanishing cosmological
constant \cite{Deser:1983mm,Higuchi:1986wu,Deser:2001us}
(see also \cite{Deser:1983tm, Higuchi:1986py,Higuchi:1989gz,Deser:2001wx, Zinoviev:2001dt,  Deser:2001pe}).
They appear as fields whose mass take special values for which
the corresponding action acquires a gauge symmetry of higher-derivative type, and hence propagate
an intermediate number of degrees of freedom between those
of a genuine massless field (subject to single-derivative
gauge invariance), and a genuine massive field (without
any gauge symmetry).

Partially-massless fields are unitary in de Sitter space
and may have phenomenological applications (see e.g.
\cite{Baumann:2017jvh, Goon:2018fyu} and references therein).
Despite being non-unitary around anti-de Sitter spacetime,
partially-massless fields are nevertheless of interest\footnote{Note that
partially-massless fields have also been of interest recently
in in the context of inflation \cite{Baumann:2017jvh, Goon:2018fyu}.},
if only because they are dual to partially-conserved
currents, that is, currents which are annihilated after
taking several divergences \cite{Dolan:2001ih}. These kinds
of currents naturally appear in free conformal field theories
of higher-derivative scalar fields, i.e. scalar fields subject
to polywave equations of the type $\Box^\ell\phi=0$,
with $\ell>1$ \cite{Bekaert:2013zya}, which are known to describe
special RG fixed points called `multi-critical isotropic
Lifshitz points' \cite{Diehl:2002ri}. The holographic dual
of this theory would be a theory of both massless and
partially-massless fields of arbitrary spin in anti-de Sitter
space, which has been studied in \cite{Bekaert:2013zya,Alkalaev:2014nsa, Brust:2016zns}
(see also \cite{Eastwood2008, Gover2009, Michel2014, Joung:2015jza}
for works on the corresponding higher spin algebras), but not worked out
in full details yet. One reason is that holographic duals of vector models feature severe nonlocalities that invalidate the usual field theory methods to construct them \cite{Maldacena:2015iua, Bekaert:2015tva, Sleight:2017pcz, Ponomarev:2017qab}.

Nevertheless, cubic interactions for partially-massless fields
of any spins have been studied
\cite{Joung:2012hz, Boulanger:2012dx, Joung:2012rv}, but complete interacting theories
featuring partially-fields in the spectrum are still lacking.
Particular attention has been given to the problem of finding
gravitational interactions and constructing what one might want
to call a theory of partially-massless gravity, i.e. an interacting
theory of a massless and a partially-massless spin-$2$ field.
Unfortunately, the search for such a non-linear theory led
to several no-go theorems, whether it is in relation with massive
and/or bimetric gravity
\cite{Hassan:2012gz,deRham:2013wv,Garcia-Saenz:2015mqi,Apolo:2016ort,Apolo:2016vkn},
with conformal gravity \cite{Deser:2012euu, Deser:2012qg},
or on general grounds \cite{Joung:2014aba, Joung:2019wbl}.
A notable exception is the recent work \cite{Boulanger:2019zic},
wherein an interacting theory of a multiplet of spin-$2$ partially-massless
fields has been proposed.

All of the aforementioned results were obtained by working
with symmetric rank-$s$ tensors to describe partially-massless
fields of spin-$s$. In this paper, we  introduce a new description
of partially-massless fields in $4d$, inspired by twistor theory
and the description of massless fields given in \cite{Hitchin:1980hp,Krasnov:2021nsq},
based on a pair of a $1$-form and a $0$-form which are also
$SL(2,\mathbb C)$ spin-tensors (see also \cite{Krasnov:2011pp,Krasnov:2016emc,Krasnov:1970bpz}
for a pure connection formulation of gravity, which is closely
related). In terms of these new field variables,
the free action for partially-massless takes a fairly simple form,
and more importantly, one can construct complete interacting theories
featuring partially-massless fields. We will illustrate this last fact
by spelling out a partially-massless higher-spin extension of self-dual
Yang--Mills, which is a generalisation of the higher-spin
extension discussed in \cite{Krasnov:2021nsq}, and a theory featuring
current interactions between a couple of massless fields
with a partially-massless one, which is complete at the cubic order.

The organisation of this paper is as follows: in Section \ref{sec:free},
we briefly recall the metric- and frame- like description 
of free partially-massless fields before introducing a new description
based on two-component spin-tensors, in Section \ref{sec:interactions}
we present two simple examples of fully interacting theories
featuring partially-massless fields, and we end up by some
concluding remarks in Section \ref{sec:disco}.

\section{Free partially-massless fields, old and new}
\label{sec:free}
\paragraph{Metric-like approach.}
Free fields are known to be in one-to-one with irreducible
representations of the spacetime  isometry group.
For de Sitter (dS) space in $(d+1)$-dimensions,
the isometry algebra is $\mathfrak{so}(1,d+1)$, whereas 
for anti-de Sitter (AdS) space in $(d+1)$-dimensions, it is
$\mathfrak{so}(2,d)$. We will hereafter denote these algebras
collectively by $\mathfrak{g}_\Lambda$.
One new feature of the representation theory
of (anti-)de Sitter algebras, as compared to that of
the Poincar\'e algebra,
is that they admit irreducible representations which are realized
as fields propagating
an intermediate number of degrees of freedom
between that of a massless field and that of a massive one, 
for a fixed value of the spin \cite{Deser:1983mm,Higuchi:1986wu,Deser:2001us}.
Consequently, these fields are called partially-massless (PM).
A spin-$s$ partially-massless field
of depth $t$, with $1 \leq t \leq s$,
can be represented by a rank-$s$ symmetric tensor
$\Phi^{a_1 \dots a_s} \equiv \Phi^{a(s)}$ 
that is subject to\footnote{In trying to save letters we abbreviate
a group of symmetric indices $a_1 \dots a_s$ as $a(s)$ and, more generally,
denote all indices do be symmetrized by the same letter.}
\begin{align}\label{masslessdict}
    & \delta_\xi\Phi^{a(s)}
    = \underbrace{\nabla^a \dots \nabla^a}_{t\ \text{times}} \xi^{a(s-t)}
    + \dots\,,
\end{align}
where the dots denote lower order derivatives terms.
In other words, the depth of a partially-massless field 
is nothing but the number of derivatives in its gauge transformation,
and the massless case corresponds to $t=1$ in our convention.
Omitting the transversality and tracelessness constraints
for $\Phi$ and $\xi$, the equations of motion reduce to 
\begin{align}
    (\square -m^2)\,\Phi^{a(s)} & = 0\,,
    && m^2 = -\Lambda\,\big((d+s-t-1)(s-t-1)-s\big)\,.
\end{align}
where, as for the massless case, the mass-like term is
proportional to the cosmological constant and depends on
the spin $s$, depth $t$ and spacetime dimension $d+1$.
The mass-like term is fixed by the gauge symmetry. While equations
of motion are simple, the action requires an intricate pattern
of auxiliary fields\footnote{This is due to the fact that
partially-massless fields are closer to the massive ones.
For a massive spin-$s$ field one has to impose transversality
on top of the Klein--Gordon equation,
which starting from $s=2$ requires auxiliary fields \cite{Fierz:1939ix}.} \cite{Zinoviev:2001dt}. 

\paragraph{Frame-like approach.}
The frame-like description of partially-massless fields was developed
in \cite{Skvortsov:2006at}, see also \cite{Khabarov:2019dvi}
for the specialization to $4d$
and \cite{Aragone:1979hx, Aragone:1980rk, Vasiliev:1980as, Lopatin:1987hz} 
for purely massless higher spin fields.\footnote{Note that
the frame-like description of fields arbitrary mixed-symmetry,
both massless and partially-massless, has been worked out, see e.g.
\cite{Boulanger:2008up, Boulanger:2008kw, Skvortsov:2009zu, Skvortsov:2009nv,Alkalaev:2009vm, Alkalaev:2011zv}.}
The key idea is to consider a (generalized) connection
of the (anti-)de Sitter algebra $W^{\mathbb{Y}}$, i.e. a one-form
that takes values in a finite-dimensional representation $\mathbb{Y}$
of the algebra that is not necessarily the adjoint one.
The simplest case is the adjoint itself, $\parbox{10pt}{\gyoung(;,;)}$ \cite{Stelle:1979aj},
for which the connection $W^{A,B}$ contains\footnote{Indices $A,B,...=0,...,d+1$ are of $\mathfrak{g}_\Lambda$ and we can decompose them as $A={a,\bullet}$, where indices $a,b,c,...$ are of the Lorentz algebra. } two one-forms valued
in finite-dimensional representations of the Lorentz subalgebra $\mathfrak{so}(1,d)$,
namely the vielbein $e^a=W^{a,\bullet}$ and the spin-connection
$\omega^{a,b}=W^{a,b}$.
In order to describe a spin-$s$ depth-$t$ partially-massless field,
one should
consider a $1$-form $W$ taking values in the finite-dimensional
irreducible representation
$\mathbb Y_{s,t}=\parbox{30pt}{\tiny\gyoung(_5{s-1},_4{s-t})}$\,.
Upon decomposing it with respect to the Lorentz algebra,
one gets a lot of auxiliary fields,
\begin{equation}
    W^{\mathbb Y_{s,t}} =\{\omega^{a(s-k),b(s-m)}\}\,,
    \qquad\text{with}\qquad
    k \in \{1,2,\dots,t\}\,,
    \quad m \in \{t,t+1,\dots,s\}\,.
\end{equation}
It is easy to construct a gauge-invariant curvature $R$ for $W$,
namely one simply defines it to be
\begin{equation}
    R[W] = \nabla W + e^a\wedge\rho(P_a)\,W\,,
\end{equation}
where $\rho$ is the representation $\mathbb Y$ of the (anti-)de Sitter
algebra.\footnote{This expression can be
thought of as originating from the curvature $F[A]=dA+\tfrac12[A,A]$
of a connection $A$ taking values in the algebra
$\mathfrak{g}_\Lambda \inplus_\rho \mathbb Y$, which is
the semi-direct sum of the (anti-)de Sitter algebra
$\mathfrak{g}_\Lambda$ with the representation $\mathbb Y$,
considered as an Abelian subalgebra. The component of
this curvature taking values in $\mathfrak{g}_\Lambda$
is the usual curvature of the (A)dS algebra, and is assumed
to vanish here, while the component in $\mathbb{Y}$ reproduces
the above formula.} This curvature is
invariant under the gauge transformations generated by
a $0$-form $\xi$ valued in the same representation $\mathbb{Y}$,
\begin{equation}\label{eq:gauge_frame}
    \delta_\xi W = \nabla\xi + e^a\,\rho(P_a)\,\xi\,,
\end{equation}
on an (anti-)de Sitter background, i.e. defined by a vielbein
$e^a$ and spin-connection $\varpi^{a,b}$ obeying
\begin{equation}
    \nabla e^a = 0\,,
    \qquad \qquad 
    R^{ab}-e^{[a} \wedge e^{b]} = 0\,,
\end{equation}
where $\nabla$ is the covariant derivative induced by $\varpi$
and $R^{a,b} = d\varpi^{a,b} + \varpi^a{}_c\wedge\varpi^{c,b}$
is its usual Lorentz curvature $2$-form. 
Note in particular that the second piece of this gauge transformations,
the one generated by the action of the transvection generators,
is algebraic (it is given by symmetrization and contraction 
of the background vielbein with the gauge parameters, and 
does not involve any derivatives).

For instance, a partially-massless spin-$2$ field is described
in this language by a connection, taking values in
$\mathbb{Y}=\parbox{10pt}{\tiny\gyoung(;)}$,
the fundamental (or vector) representation of the (anti-)de Sitter
algebra $\mathfrak{g}_\Lambda$. Such a connection has components
$W^{\gyoung(;)}=\{w^a,w\}$, i.e. it is composed of two $1$-forms,
valued in the vector and scalar representation of
the Lorentz algebra respectively. Their curvature simply read
\begin{equation}
    R^a = \nabla w^a + e^a \wedge w\,,
    \qquad 
    R = \nabla w - e^a \wedge w_a\,,
\end{equation}
while the gauge transformations are given by
\begin{equation}
    \delta_{\xi,\epsilon} w^a = \nabla\xi^a + e^a\,\epsilon\,,
    \qquad 
    \delta_{\xi,\epsilon} w = \nabla\epsilon - e^a\,\xi_a\,,
\end{equation}
where $\xi^a$ and $\epsilon$ are the two $0$-form gauge parameters.
Let us briefly review how one can recover the metric-like formulation
discussed previously \cite[Sec. 5.1]{Skvortsov:2006at}. First,
note that one can gauge-fix to zero the component $w$ upon using
its gauge symmetry generated by $\xi^a$. The residual gauge
transformations (i.e. which preserve the gauge choice $w=0$)
are those generated by $\epsilon$ and $\xi_a = -\nabla_a\epsilon$,
i.e.
\begin{equation}
    \delta_\epsilon w_{a|b} = -\nabla_a\nabla_b\,\epsilon
    + \eta_{ab}\,\epsilon\,,
\end{equation}
where $w_{b|a} = e_b^\mu\,w_\mu^c\,\eta_{ac}$. Imposing that
the curvature $R$ of $w$ vanishes in the gauge $w=0$ implies
that the antisymmetric part of $w_{a|b}$ vanishes,
\begin{equation}
    R\rvert_{w=0} = 0
    \qquad \Rightarrow \qquad 
    w_{[a|b]} = 0\,.
\end{equation}
This is a first sign that one can recover the symmetric
rank-$2$ tensor subject to a two-derivative gauge transformation,
which encodes the PM spin-$2$ field in the metric-like formulation,
as the symmetric part of the $1$-form $w^a$.
Inspecting the Bianchi identities for the curvature $R^a$,
one finds that its only possible non-trivial component 
is encoded by a hook, so that one can impose
\begin{equation}
    R^a = e_b\wedge e_c\,C^{ab,c}\,,
\end{equation}
where $C^{ab,c}$ is a $0$-form which takes values 
in the irrep $\gyoung(;;,;)$ of the Lorentz algebra.
The above example is representative of the frame-like
description of partially-massless field: for a spin-$s$
and depth-$t$ field, one can impose the zero-curvature
equations
\begin{equation}
    R^{a(s-m),b(s-n)} = 0\,,
    \qquad 
    m\neq 1 \quad\text{and}\quad n \neq t\,,
\end{equation}
and 
\begin{equation}
    R^{a(s-1),b(s-t)} = C^{a(s-1)c,b(s-t)d}\,e_c\wedge e_d\,,
\end{equation}
where $C$ is a $0$-form, that can be thought of as
a partially-massless version of the Weyl tensor. 
The metric-like partially-massless field can be found
in the connection $e^{a(s-1)}$ valued in the totally symmetric
irrep of the Lorentz algebra, and the above zero-curvature equations
expresses the intermediate/auxiliary connections $\omega^{a(s-1),b(m)}$
with $m=1,\dots,s-t-1$ as $m$ derivatives of the PM field, 
while the last equation equates the $0$-form $C$ to a particular
traceless projection of $s-t+1$ derivatives of the PM field.

One can build a gauge-invariant action from the above curvature,
however, this action exhibits an intricate pattern involving
the `auxiliary connections' \cite{Skvortsov:2006at}.
Let us specialize this construction to $4d$, where it is advantageous
to use the two-component spinor language.

\paragraph{Twistor-inspired/chiral approach.}
The advent of twistor theory lead to a new geometrical
understanding of massless fields in $4d$ in terms of holomorphic
structures on a $3d$ complex manifold that is twistor space
\cite{Eastwood:1981jy, Atiyah:1979iu, Hitchin:1980hp} (see also
the textbooks \cite{Huggett:1986fs, Penrose:1986ca, Ward:1990vs, Mason:1991rf} and, for instance, the recent review
\cite{Adamo:2017qyl}). Although we will not use directly twistor
theory in our description of partially-massless fields,
it is very much inspired by it, and is a straightforward
extension of the approach proposed for massless fields
in \cite{Hitchin:1980hp, Krasnov:2021nsq}.

At the algebraic level, this relies on the low dimensional
isomorphism $\mathfrak{sl}(2,\mathbb C) \cong \mathfrak{so}(1,3)$.
The latter relates a Lorentz vector $V^a$ to 
a $\mathfrak{sl}(2,\mathbb C)$-bi-spinor $V^{AA'}$,
where both $A=1,2$ and $A'=1,2$ are two-component spinor indices.
More generally, finite-dimensional irreducible
representations of $\mathfrak{so}(1,3)$, which are
mixed-symmetric traceless tensor $T^{a(m),b(n)}$
correspond to a spin-tensor carrying two groups of $m+n$ and $m-n$ totally symmetrized (un)primed indices,
\begin{equation}
    T^{a(m),b(n)}
    \qquad\longleftrightarrow\qquad
    \big(T^{A(m+n),A'(m-n)},\,\, T^{A(m-n),A'(m+n)}\big)\,.
\end{equation}
As usual, in the Lorentzian signature the two spin-tensors
are complex conjugate of each other. In the Euclidian
or split signature, they are independent real spin-tensors.  
Unprimed spinor indices are raised and lowered
with the invariant tensor $\epsilon_{AB}$ and its inverse
$\epsilon^{AB}$, in the sense that 
$\epsilon^{AC}\,\epsilon_{BC} = \delta^A_B$, via
\begin{equation}
    \xi^A = \epsilon^{AB}\,\xi_B\,,
    \qquad\qquad
    \xi_B = \xi^A\,\epsilon_{AB}\,,
\end{equation}
and similarly for primed indices. In this two-component
spinor language, the $\mathfrak{g}_\Lambda$-connection
consists of a vierbein $e^{AA'}$, the self-dual part of
the spin-connection $\omega^{AA}$, and its anti-self-dual
part $\omega^{A'A'}$. The zero-curvature equations 
for this connection are given by
\begin{equation}
    R_{AA} = H_{AA}\,,
    \qquad
    R_{A'A'} = H_{A'A'}\,,
    \qquad
    \nabla e_{AA'} = 0\,,
\end{equation}
with
\begin{equation}
    R_{AA} := d\omega_{AA}
    + \omega_{AB}\wedge\omega\fud{B}{A}\,,
    \qquad 
    R_{A'A'} := d\omega_{A'A'}
    + \omega_{A'B'}\wedge\omega\fud{B'}{A'}\,,
\end{equation}
are the self-dual and anti-self-dual parts
of the Lorentz curvature $2$-form,
and where we introduced the two-forms
\begin{equation}
    H_{AA} := e_{AB'} \wedge e\fdu{A}{B'}\,,
    \qquad
    H_{A'A'} := e_{BA'} \wedge e\fud{B}{A'}\,,
\end{equation}
which define a basis of self-dual and
anti-self-dual $2$-forms respectively. There is also
the $3$-form basis, defined as
\begin{equation}\label{eq:basis_3-forms}
    \hat e_{AA'} := H_{AB} \wedge e\fud{B}{A'}\,.
\end{equation}
In particular, the $2$-forms $H_{AA}$ and $H_{A'A'}$
verify
\begin{equation}
    H_{AB} \wedge H_{A'B'} = 0\,,
\end{equation}
and the identities
\begin{equation}\label{eq:sym_prop_H}
    H_{AA} \wedge e_{AB'} = 0
    \qquad\Rightarrow\qquad 
    H_{AA} \wedge H_{AB} = 0\,,
\end{equation}
which will be useful later on (for more details,
see e.g. \cite{Krasnov:2020lku}). 

It was shown in \cite{Krasnov:2021nsq} that
for massless fields, we can take the self-dual parts
of the very `last' spin-connection (by which we mean
the component of the $\mathfrak{g}_\Lambda$-connection
valued in the `biggest' Lorentz Young diagram, that is,
the Young diagram with the same shape as the one labelling
the $\mathfrak{g}_\Lambda$-irrep) and of the Weyl tensor
as our dynamical variables. Indeed, we will show that this leads to
a simple action. In tensor language, the last spin-connection
for a spin-$s$ and depth-$t$ partially-massless field
is a one-form $\omega^{a(s-1),b(s-t)}$ and the Weyl tensor
is of the form $C^{a(s),b(s-t+1)}$, where the indices merely
indicate the symmetry type of a tensor. In the spinorial language,
the self-dual components of these two fields are thus 
\begin{align}
    \omega^{A(2s-t-1),A'(t-1)}
    &&& \Psi^{A(2s-t+1),A'(t-1)}\,,
\end{align}
and their anti-self-dual cousins can be obtained via $t \to 2s-t$
for $\omega$ and $t \to 2s-t+2$ for $\Psi$.
The chiral approach deals with one pair of such fields
and ignores the duals thereof. 
\begin{figure}[!ht]
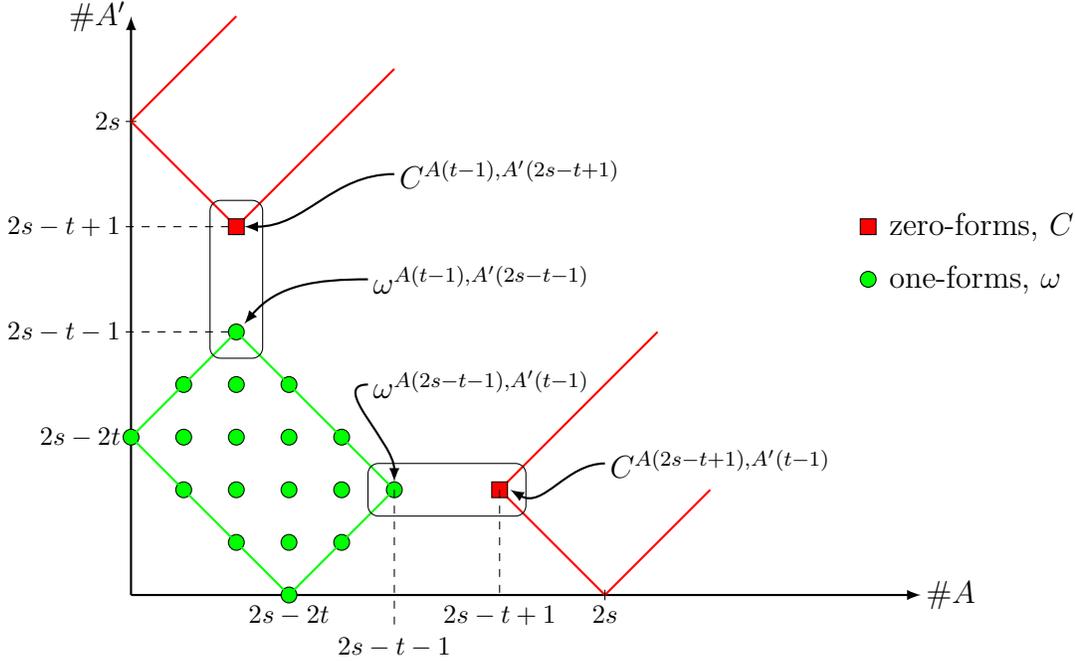

    \FIELDS
    \label{fig:fields}
    \caption{A diagram to show fields/coordinates involved
    into the description of partially-massless higher spin
    fields. Along the horizontal/vertical axe, we have
    the number of unprimed/primed indices on a spin-tensor. Components of the $1$-form connection are represented
    by green circles, while the $0$-forms (the Weyl tensor
    and its descendants) are represented by red rectangles. By descendants we mean the on-shell nontrivial derivatives of the Weyl tensor, which are associated with the coordinates on the on-shell jet space \cite{Penrose:1986ca}.  }
\end{figure}
In particular, for the spin-$2$ field of depth $t=2$,
i.e. a partially-massless graviton, we find $\omega^{A,A'}$
and $\Psi^{A(3),A'}$. 
In the spin-$s$, case the decomposition of $\omega$ into
irreducible spin-tensors reads
\begin{equation}
    \begin{aligned}\label{eq:irreducible_decompo}
        \omega^{A(2s-t-1),A'(t-1)} &
        = e\fud{A}{B'}\Phi^{A(2s-t-2),A'(t-1)B'}
        + e\fdu{B}{A'}\Phi^{A(2s-t-1)B,A'(t-2)} \\
        & \qquad +e_{BB'} \Phi^{A(2s-t-1)B,A'(t-1)B'}
        + e^{AA'}\Phi^{A(2s-t-2),A'(t-2)}\,,
    \end{aligned}
\end{equation}
where $\Phi$ are $0$-forms.
Two of these components are unphysical and can be gauged away,
since the gauge transformation of $\omega$ reads
\begin{equation}\label{eq:gauge_transformations}
    \begin{aligned}
        \delta_{\xi,\eta}\omega^{A(2s-t-1),A'(t-1)}
        & = \nabla \xi^{A(2s-t-1),A'(t-1)} \\
        & \qquad\qquad + e^{AA'}\eta^{A(2s-t-2),A'(t-2)}
        + e\fud{A}{B'}\eta^{A(2s-t-2),A'(t-1)B'}\,,
    \end{aligned}
\end{equation}
and contains both a differential part (the first term),
and an algebraic part (the second and third terms). The latter,
hereafter referred to as a {\it shift symmetry},
can therefore be used to gauge away the first and fourth terms
in the irreducible decomposition \eqref{eq:irreducible_decompo}.
After this gauge fixing, the connection $\omega$ is given by
\begin{align}
    \omega^{A(2s-t-1),A'(t-1)}&= e\fdu{B}{A'}\Phi^{A(2s-t-1)B,A'(t-2)}
    +e_{BB'} \Phi^{A(2s-t-1)B,A'(t-1)B'}\,,
\end{align}
and is subject to the residual gauge symmetry
\begin{subequations}
    \begin{align}
        \delta \Phi^{A(2s-t),A'(t-2)}
        & = \nabla\fud{A}{B'} \xi^{A(2s-t-1),A'(t-2)B'}\,,\\
        \delta \Phi^{A(2s-t),A'(t)} & = \nabla^{AA'} \xi^{A(2s-t-1),A'(t-1)}\,,
    \end{align}
\end{subequations}
expressed in terms of its two irreducible components. 
Note that the gauge symmetry \eqref{eq:gauge_transformations}
is nothing but the two-component spinor translation of
the gauge symmetry \eqref{eq:gauge_frame} in the frame-like
approach, and in particular, the shift symmetry here is
simply the algebraic part of the gauge symmetry of the `last
connection'.

\paragraph{Action.}
We propose the following action 
\begin{align}
   {S_{s,t}[\omega,\Psi]
   =\int \Psi^{A(2s-t+1),A'(t-1)}\,H_{AA}
   \wedge \nabla \omega_{A(2s-t-1),A'(t-1)}\,,}
   \label{eq:free_PM_action}
\end{align}
for the description of a spin-$s$ partially-massless field
of depth-$t$. This action is invariant under the gauge symmetries
\eqref{eq:gauge_transformations} thanks to the property
\eqref{eq:sym_prop_H} of the background. Notice also that
this action is of presymplectic AKSZ-type \cite{Alkalaev:2013hta},
which is not that surprising considering that the frame-like action
for Gravity \cite{Grigoriev:2020xec} and Conformal/Weyl Gravity
\cite{Dneprov:2022jyn} are also of this type, and that the relevance
of this approach for higher-spin theories is established
\cite{Sharapov:2016qne, Sharapov:2021drr}.

Another noteworthy feature of the above action is that
it is not manifestly real in the Lorentzian signature,
as is the well-known cases of (self-dual) Yang--Mills theory \cite{Chalmers:1996rq} and gravity \cite{Krasnov:2011pp,Krasnov:2016emc,Krasnov:1970bpz} that can be formulated in terms of chiral field variables. Nevertheless, it is worth mentioning that the use of chiral field variables does not imply that the theory is actually chiral (parity-violating) or non-unitary. This is always true for free theories that have the same degrees of freedom as their non-chiral relatives. The free action of \cite{Chalmers:1996rq} corresponds to $s=1$, $t=1$ of \eqref{eq:free_PM_action}.

The equations of motion obtained from \eqref{eq:free_PM_action} are 
\begin{align}\label{eq:free_EOM}
    H_{AA} \nabla \omega_{A(2s-t-1),A'(t-1)} & = 0\,, &
    H_{AA} \nabla\Psi^{A(2s-t+1),A'(t-1)} & = 0\,.
\end{align}
There are two noteworthy cases: $t=1$ which corresponds to
massless fields, and in which case the above action reproduces
the one proposed in \cite{Krasnov:2021nsq}, and $t=s$,
which corresponds to maximal depth partially-massless fields,
and for which the spin-connection is balanced (meaning it has the same number of primed and unprimed indices, as opposed
to the massless case where it is completely unbalanced).

These equations can be taken as a starting point
to build a free differential algebra (FDA) formulation of
partially-massless fields, see  \cite{Skvortsov:2006at,Skvortsov:2009zu,Skvortsov:2009nv,Ponomarev:2010st,Alkalaev:2011zv,Khabarov:2019dvi}. Indeed, they can be read as expressing 
the fact that the first derivatives of $\omega$ and $\Psi$
are in the kernel of an operator determined by the background
self-dual $2$-form $H_{AA}$ (symmetrization for $\omega$,
contraction for $\Psi$). These operators are nothing but
components of the presymplectic
form used to build the action \eqref{eq:free_PM_action}.
The FDA is obtained by parametrizing $\nabla\omega$
and $\nabla\Psi$ as the most general elements in the kernel
of this presymplectic form, i.e.
\begin{subequations}\label{firststepsFDA}
    \begin{align}
        \nabla\omega_{A(2s-t-1),A'(t-1)}
        & = e\fdu{A}{B'}\,\omega_{A(2s-t-2),A'(t-1)B'}
        + e_{AA'}\,\omega_{A(2s-t-2),A'(t-2)}\,, \\
        \nabla\Psi_{A(2s-t+1),A'(t-1)}
        & = e\fud{B}{A'}\,\Psi_{A(2s-t+1)B,A'(t-2)}
        + e^{BB'}\,\Psi_{A(2s-t+1)B,A'(t-1)B'}\,,
    \end{align}
\end{subequations}
and imposing that the resulting equations
are integrable. Typically, this condition leads to constraints
on the first derivatives of the components of the elements
in the kernel of the symplectic form, and one should repeat
the procedure (i.e. find the most general form of the first
derivatives of these new fields compatible with integrability,
thereby introducing new fields, and imposing once more
the integrability of this equation, etc \dots). See e.g.
\cite{Alkalaev:2013hta} or \cite[Sec. 4]{Sharapov:2021drr}
for a review. The outcome of this procedure is to build two modules
of the (A)dS algebra $\mathfrak{g}_\Lambda$:
\begin{itemize}
\item A finite-dimensional one, which is spanned by the $1$-forms
$\omega^{A(2s-m-n),A'(n-m)}$ and their complex conjugate,
with $1 \leq m \leq t$ and $t \leq n \leq s$. This corresponds
to the $\mathfrak{g}_\Lambda$-module ${\tiny\gyoung(_5{s-1},_4{s-t})}$
used in the frame-like formulation;
\item An infinite-dimensional one, spanned by the $0$-forms
$\Psi^{A(2s-t+m+n),A'(t-m+n)}$ with $n\geq0$ and $1 \leq m \leq t$,
which corresponds to the derivatives of the self-dual Weyl tensors
unconstrained by equations of motion or Bianchi identities.
\end{itemize}
The pattern of connections, and descendants of the Weyl tensor
is illustrated in Figure \ref{fig:fields} and was already
detailed in \cite{Skvortsov:2006at}, see also \cite{Boulanger:2008up, Boulanger:2008kw, Skvortsov:2009zu, Skvortsov:2009nv,Alkalaev:2009vm, Alkalaev:2011zv, Ponomarev:2010st}.

Let us dwell a little on the maximal depth case $t=s$.
In vector language, the last connection decomposes as
\begin{equation}
    \Yboxdim{10pt}
    \omega^{a(s-1)} \simeq {\footnotesize \gyoung(_5{s})
    \oplus \gyoung(_5{s-1},;) \oplus \gyoung(_4{s-2})}
\end{equation}
under the Lorentz group, and is subject to the algebraic symmetry
\begin{equation}
    \delta_\epsilon \omega^{a(s-1)} = e^{\{a}\,\epsilon^{a(s-2)\}}\,,
\end{equation}
where $\{\dots\}$ denotes the traceless projection
of symmetrized indices. This algebraic symmetry removes
the trace part ${\tiny\gyoung(_4{s-2})}$ in the irreducible
decomposition of $\omega^{a(s-1)}$. It may however be surprising
at first glance that in the two-component spinor language,
one has two parameters for the algebraic symmetry of $\omega$,
namely $\eta^{A(s-2),A'(s-2)}$ and $\eta^{A(s-2),A'(s)}$.
The first one simply corresponds to $\epsilon$, converted
in spinor language, but the second one appears to have no counterpart
in the vector language. This is not accidental: in fact,
this additional parameter has the same symmetry has the 
anti-self-dual part of the hook component of $\omega$,
and its r\^ole is simply to remove it. This is consistent
with the fact that, in spinor language, $\omega$ has two
irreducible components, corresponding respectively to 
symmetric rank-$s$ tensor and the self-dual part of a hook
tensor, and is also in accordance with the counting of degrees
of freedom detailed below. Such additional symmetry is also present in the FDA form \cite{Ponomarev:2010st,Khabarov:2019dvi} of Zinoviev's description of partially-massless fields \cite{Zinoviev:2001dt,Zinoviev:2008ze}. 

\begin{figure}[!ht]
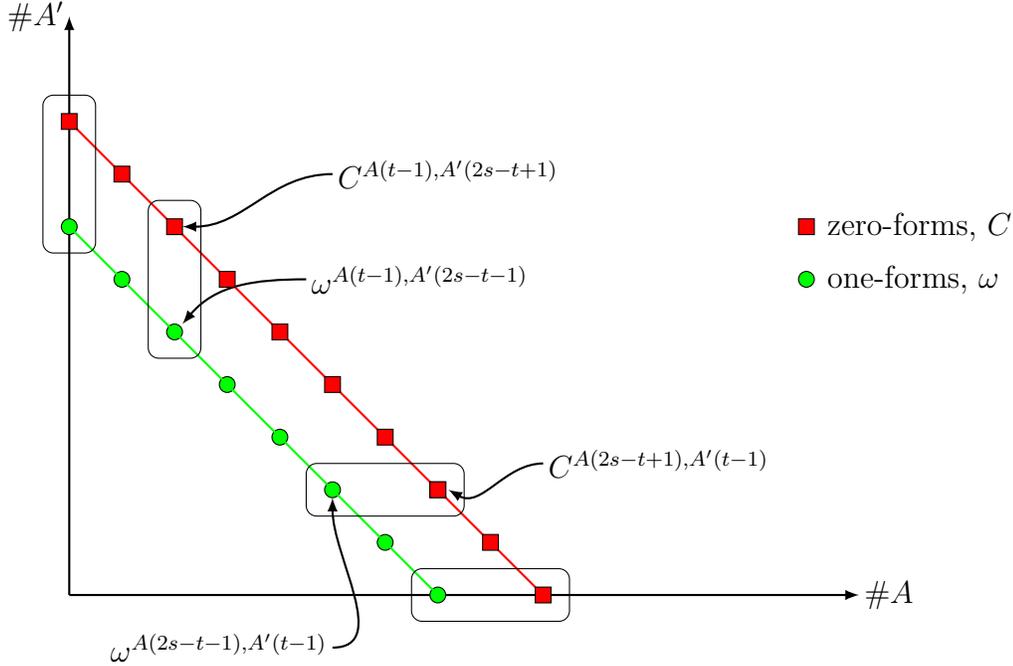

    \FIELDSDescr
    \label{fig:scale}
    \caption{For a given spin-$s$, the fields grouped horizontally/vertically
    correspond to chiral/anti-chiral description of depth-$t$
    partially-massless fields. There are two descriptions for each admissible $s$, and $t$. The group on each of the axes describes
    massless fields in terms of (anti-)chiral variables. It is clear that extrapolation of one description
    beyond $t>s$ does give the other one.}
\end{figure}

Massless spinning fields, described as in \cite{Hitchin:1980hp,Krasnov:2021nsq},
can propagate on self-dual backgrounds. This is due to the fact that
the fields $\Psi^{A(2s)}$ and $\omega^{A(2s-2)}$
do not have any primed indices, hence, $\nabla^2 \xi^{A(2s-2)}\equiv0$
on a self-dual background, which ensures the gauge invariance
of the action. However, partially-massless fields are always
described by mixed spin-tensors, i.e. have both primed and
unprimed indices. 
The action \eqref{eq:free_PM_action} as well as the equations
of motion \eqref{eq:free_EOM} remain consistent in Minkowski space,
the difference being that the corresponding solution space
is not an irreducible representation of the Poincar\'e group
(see e.g. \cite{Brink:2000ag, Boulanger:2008up, Boulanger:2008kw, Alkalaev:2009vm, Alkalaev:2011zv}). 

\paragraph{Degrees of freedom.}
Let us justify the main claim of the previous paragraphs,
which is that the action \eqref{eq:free_PM_action} does describe
a partially-massless spin-$s$ and depth-$t$ field in $4d$.
To do so, we will show that the solutions of the resulting
equations of motion propagate the correct number of degrees of freedom,
namely $2t$ (irrespectively of the spin).
In our case, the equations of motion are first order differential
equations for the fields $\Psi$ and $\omega$.
The number of physical degrees of freedom 
propagated by an arbitrary field, which is a solution of
an involutive system of equations,
is given by the formula \cite{Kaparulin:2012px}
\begin{equation}
    N_{dof} = \tfrac12\,\sum ^{\infty }_{k=0}
    k\left(e_k-i_k-g_k\right)\,,
\end{equation}
where $e_k$ is the number of equations of order $k$ in the system,
$i_k$ number of (gauge) identities of $k$-th order, and $g_k$ is
the number of gauge symmetry generators of order $k$ (here,
the order is the number of derivatives).
Let us recall that an involutive system of order $n$ is defined
in \cite{Kaparulin:2012px} as a system of equations such that
any differential consequence of these equations, of order $n$
or less, is already a part of the system.
In our case, the equation of motion for the field $\Psi$ is given by,
\begin{align}
    H_{AA} \wedge \nabla\Psi^{A(2s-t+1),A'(t-1)}
    \propto \widehat{e}_{AB'}\nabla\fdu{A}{B'}\Psi^{A(2s-t+1),A'(t-1)}=0\,,
\end{align}
where $\hat e_{AB'}$ are the basis $3$-forms introduced
in \eqref{eq:basis_3-forms} above. 
Using it, we can write down the set of independent equations
of motion as
\begin{equation}
    E^{A(2s-t),A'(t-1)|B'}
    = \nabla\fdu{B}{B'}\Psi^{BA(2s-t),A'(t-1)}=0\,,
\end{equation}
and easily count that these are $e_1=2t(2s-t+1)$ equations
of first order.
The field $\Psi$ does not have any gauge symmetry, 
hence $g_k=0$ for all $k$. Now since the field $\omega$ has a first
order gauge symmetry, the $\Psi$-field after integrating by parts in the action, satisfy the Bianchi identity of second order. 
Explicitly, this identity is given by,
\begin{align}
    \nabla_{FF'} E^{FA(2s-t-1),F'A'(t-1)} = 0\,,
\end{align}
which consists in $i_2=t\,(2s-t)$ identities of the second order. 
Thus, the number of  physical degrees of freedom described by
the field $\Psi$ is
\begin{equation}
    N_{dof}(\Psi)=\tfrac12\,\big[2t(2s-t+1)-2t(2s-t)\big]=t\,.
\end{equation}
Similarly, the equations of motion for the field $\omega$
read
\begin{align}
    H_{AA}\wedge \nabla\omega_{A(2s-t-1),A'(t-1)}=H_{AA}\wedge e_{DD'}\nabla^{DD'}\omega_{A(2s-t-1),A'(t-1)}=0\,,
\end{align}
and, upon using the decomposition of $\omega$ into
its irreducible components,
\begin{equation}
    \begin{aligned}
        \omega^{A(2s-t-1),A'(t-1)} 
        & = e\fud{A}{B'}\Phi^{A(2s-t-2),A'(t-1)B'}
        +e\fdu{B}{A'}\Phi^{A(2s-t-1)B,A'(t-2)} \\
        & \qquad + e_{BB'} \Phi^{A(2s-t-1)B,A'(t-1)B'}
        +e^{AA'}\Phi^{A(2s-t-2),A'(t-2)}\,,
    \end{aligned}
\end{equation}
takes the form
\begin{align}
    \nabla\fdu{A}{F'}\Phi_{A(2s-t),F'A'(t-1)}
    + \nabla_{AA'}\Phi_{A(2s-t),A'(t-2)} = 0\,.
\end{align}
These are $e_1=t\,(2s-t+2)$ equations of first order.
The gauge transformations
are of first order, and generated by $g_1=t\,(2s-t)$ parameters. 
Since there are no additional identities, the number 
of degrees of freedom propagated by $\omega$ is
\begin{equation}
    N_{dof}(\omega) = \tfrac12\,\big[(2s-t+2)t-(2s-t)t\big]=t\,,
\end{equation}
and hence $\Psi$ and $\omega$ contain, in total, $2t$ physical
degrees of freedom. 
In particular, for massless field ($t=1$), we recover
$2$ degrees of freedom, as expected, while for
the partially-massless graviton ($t=2$), we find $4$ degrees
of freedom, in conformity with expectations.\footnote{The same counting of degrees of freedom is suggested by the first step \eqref{firststepsFDA} towards the FDA form of the equations.}

Note that the counting of degrees of freedom presented
here applies for any values of $t$. In particular, when $t>s$,
we see that the number of degrees of freedom keeps increasing
and is larger than the one expected for a spin-$s$ field
of any depth. This is another indication that, despite the fact
that the pairs of fields $(\omega,\Psi)$ can still be considered
for $t>s$, and the action \eqref{eq:free_PM_action} still
makes sense, their interpretation remains elusive and should not
be related to PM fields (our proposal is that it gives two massive fields, see Appendix \ref{app:beyond}).

\section{Interactions}
\label{sec:interactions}
Since we have a well-defined free action, the next task
is to look for interacting theories. In this section,
we will consider two simple types of possible interactions
using the new description presented in this paper. 

\subsection{Yang--Mills interactions}
\label{sec:YM}
First, we will consider Yang--Mills interactions
for partially-massless fields, which are straightforward
generalization of the higher spin extension of self-dual
Yang--Mills theory introduced in \cite{Krasnov:2021nsq},
and recently revisited in \cite{Tran:2021ukl,Herfray:2022prf, Adamo:2022lah,Tran:2022tft}, see also \cite{Steinacker:2022jjv}.
This type of interaction is obtained by first extending
the spin-connection $\omega^{A(2s-t-1),A'(t-1)}$
and the Weyl tensor $\Psi^{A(2s-t+1),A'(t-1)}$ of
a partially-massless spin-$s$ and depth-$t$ field to take
values in a Lie algebra $\mathfrak{g}$ equipped with an ad-invariant
bilinear form\footnote{Recall that a bilinear form is called ad-invariant
if it verifies $\pmb([x,y],z\pmb)=\pmb(x,[y,z]\pmb)$ for any elements
$x,y,z \in \mathfrak{g}$.} that we will denote by $\pmb(-,-\pmb)$. Next,
we can pack up together the spin-connections for partially-massless
fields of all spin and depth into a single $1$-form, 
\begin{align}
    \omega = \sum_{s=1}^\infty\sum_{t=1}^s\,\omega_{s,t}(x|y)\,,
    \qquad
    \omega_{s,t}(x|y)
    := \frac{\omega_{A(2s-t-1)A'(t-1)}}{(2s-t-1)!(t-1)!}\,
    y^A \dots y^A\,\bar{y}^{A'} \dots \bar{y}^{A'}\,,
\end{align}
whose curvature is defined by the usual formula
\begin{equation}
    F = \nabla \omega + \tfrac12\,[\omega,\omega]\,,
\end{equation}
where the bracket above should be understood as
the $\mathbb{C}[y,\bar y]$-linear extension of the Lie bracket
of the Yang--Mills algebra $\mathfrak{g}$. 
More concretely, the Lie bracket of $\omega$ with itself 
is given by
\begin{equation}
    [\omega,\omega]_{s,t}
    = \sum_{\substack{s_1+s_2=s+1\\ t_1+t_2=t+1}}\,
    [\omega_{s_1,t_1},\omega_{s_2,t_2}]\,,
\end{equation}
where the subscript $(s,t)$ denotes the component
of degree $2s-t-1$ in $y$ and $t-1$ in $\bar y$.
Packing up in a similar way the differential gauge parameters
associated with each spin-connection into a $0$-form $\xi$,
we can define an extension of the free gauge symmetry
\eqref{eq:gauge_transformations} via
\begin{equation}\label{eq:YM_gauge_transfo}
    \delta_\xi\omega = \nabla\xi+[\omega,\xi]\,,
    \qquad\qquad
    \delta_\xi\Psi = [\Psi,\xi]\,,
\end{equation}
under which the curvature transforms according to
\begin{equation}
    \delta_\xi F = \nabla^2\xi + [F,\xi]\,,
\end{equation}
where the first term can be re-written as
\begin{equation}\label{eq:curvature_gauge}
    \nabla^2\xi = (H\fdu{A}{B}\,y^A\,\partial_B
    + H\fdu{A'}{B'}\,\bar y^{A'}\,\partial_{B'})\,\xi\,.
\end{equation}
Similarly, we can pack up the shift symmetry parameters
into a single $0$-form $\eta$, and write it as
\begin{equation}
    \delta_\eta\omega = e_{AA'}\,y^A\,(\bar y^{A'}
    +\partial^{A'})\,\eta\,,
\end{equation}
so that the curvature transforms as
\begin{equation}\label{eq:shift_curvature}
    \delta_\eta F = -e_{AA'}\,y^A\,(\bar y^{A'} + \partial^{A'})\,
    (\nabla\eta+[\omega,\eta])\,,
\end{equation}
since the vierbein is torsionless and does not take values 
in the Lie algebra $\mathfrak{g}$. We will consider the action
\begin{equation}
    \begin{aligned}
        S_{PMYM}[\omega,\Psi] & = \langle\Psi \mid \tfrac12\,H_{AA}\,y^A y^A
        \wedge F \rangle\\
        & := \sum_{1 \leq t \leq s}\,
        \tfrac1{(2s-t-1)!(t-1)!}\,\int \pmb(\Psi^{A(2s-t+1),A'(t-1)},
        H_{AA} \wedge F_{A(2s-t-1),A'(t-1)}\pmb)\,,
    \end{aligned}
\end{equation}
which defines an complete interacting theory for partially-massless fields. The interactions are of the Yang--Mills type. 
This action is invariant under shift symmetry
since its variation under this transformation will produce
a term $H_{AA} \wedge e_{AB'} = 0$, as can be seen
from \eqref{eq:shift_curvature}. Its variation under the gauge
transformations \eqref{eq:YM_gauge_transfo} is given by
\begin{equation}
    \delta_\xi S_{PMYM} = \langle[\Psi,\xi] \mid
    \tfrac12\,H_{AA}\,y^A y^A \wedge F \rangle
    + \langle\Psi \mid \tfrac12\,H_{AA}\,y^A y^A \wedge
    (\nabla^2\xi+[F,\xi])\rangle = 0\,,
\end{equation}
and vanishes due to the fact that the term $\nabla^2\xi$
produces $H_{AA} \wedge H_{AB}=0=H_{AA} \wedge H_{A'B'}$
according to \eqref{eq:curvature_gauge}, and the two remaining
terms cancel one another due to the ad-invariance of the bilinear 
form on $\mathfrak{g}$.

\subsection{Current Interactions}
\label{sec:current}
Consider the functional
\begin{equation}
    S_{int}[\omega,\Psi] = \int T^{A(2s-t),A'(t)}(\Psi)\,
    \omega_{A(2s-t-1),A'(t-1)}\,\hat e_{AA'}
\end{equation}
where the spin-tensor $T^{A(2s-t+1),A'(t)}(\Psi)$ is
a $0$-form built out of Weyl tensors of some (partially-)massless
fields (of possibly different spin and depth), which verifies
\begin{equation}
    \nabla_{BB'} T^{A(2s-t)B,A'(t-1)B'}(\Psi) \approx 0\,,
\end{equation}
where the symbol $\approx$ signifies that the spin-tensor
$T(\Psi)$ is divergenceless only on-shell. This term is invariant
under the shift symmetry, as a consequence of the fact that
\begin{equation}
    e_{AA'}\wedge\hat e_{BB'} = -\tfrac14\,
    \epsilon_{AB}\,\epsilon_{A'B'}\, {\rm vol}
    \qquad\Rightarrow\qquad 
    e_{AA'}\wedge\hat e_{AB'} = 0\,,
\end{equation}
where `${\rm vol}$' denotes a volume form on the background,
and the fact that $\Psi$ is assumed to be inert under this symmetry.
Under the differential gauge symmetry, the variation of this term
reads
\begin{subequations}
    \begin{align}
        \delta_\xi S_{int}[\omega,\Psi]
        & = \int T^{A(2s-t),A'(t)}(\Psi)\,
        \nabla\xi_{A(2s-t-1),A'(t-1)}\,\hat e_{AA'} \\
        & = -\int \nabla^{BB'}\,T^{A(2s-t),A'(t)}(\Psi)\,
        \xi_{A(2s-t-1),A'(t-1)}\,e_{BB'}\,\hat e_{AA'} \\
        & = \tfrac14\,\int \nabla_{BB'} T^{A(2s-t-1)B,A'(t-1)B'}(\Psi)\,
        \xi_{A(2s-t-1),A'(t-1)}\,{\rm vol} \approx 0\,,
    \end{align}
\end{subequations}
and vanishes on-shell. It therefore provides a good starting point
to construct interactions for partially-massless fields.

Indeed, divergenceless spin-tensors are fairly easy to construct
out of the Weyl tensors of a pair of massless fields.
Consider for instance the Bel--Robinson tensor
\begin{equation}
    T_{abcd} = \tfrac14\,\left(C_{a}{}^{p}{}_{b}{}^{q}C_{cpdq}
    + *C_{a}{}^{p}{}_{b}{}^{q}*C_{cpdq}\right)\,,
\end{equation}  
where $C_{abcd}$ is the gravitational Weyl tensor
and $*$ is the Hodge dual operator, i.e. 
$*C_{abcd}=\epsilon_{ab}{}^{pq}C_{pqcd}$. This tensor is divergenceless
as a consequence of Einstein's equation in vacuum. In spinor
notations, this tensor takes an especially simple form,
namely it is given by the product of the self-dual and anti-self-dual
Weyl tensor,
\begin{equation}
    T_{A(4),A'(4)} = \Psi_{A(4)}\, \Psi_{A'(4)}\,,
\end{equation}
and suggests the generalization (see \cite{Gelfond:2006be} for a complete set of currents)
\begin{equation}\label{simplecurrents}
    T_{A(2s_1),A'(2s_2)} = \Psi_{A(2s_1)}\,\Psi_{A'(2s_2)}\,,
\end{equation}
given by the product of the Weyl tensors of two massless fields
of spin $s_1$ and $s_2$. This spin-tensor will be divergence-free
as a consequence of the equation of motion
\begin{equation}
    \nabla\fud{B}{B'}\,\Psi_{A(2s_1-1)B} \approx 0\,,
    \qquad \qquad 
    \nabla\fdu{B}{B'}\,\Psi_{A'(2s_2-1)B'} \approx 0\,,
\end{equation}
for these Weyl tensors.

We will consider the one-parameter family of actions
\begin{equation}\label{eq:current_interaction}
    S[\omega,\Psi] = S_{free}[\omega,\Psi]
    + \alpha\,S_{int}[\omega,\Psi]\,,
    \qquad 
    \alpha \in \mathbb C\,,
\end{equation}
whose first piece,
\begin{equation}
    \begin{aligned}
        S_{free}[\omega,\Psi]
        & = \int \Psi^{A(2s-t)}\,H_{AA}\wedge\nabla\omega_{A(2s-t-2)}
        + \Psi^{A'(t)}\,H_{A'A'}\wedge\nabla\omega_{A'(t-2)} \\
        & \qquad\qquad + \Psi^{A(2s-t+1),A'(t-1)}\,H_{AA}\wedge
        \nabla\omega_{A(2s-t-1),A'(t-1)}\,,
    \end{aligned}
\end{equation}
is the sum of the free actions for the massless fields
of spin $s-\tfrac{t}2$ and $\tfrac{t}2$ as well as for
the partially-massless field of spin-$s$ and depth-$t$,
and the second piece is the current interaction
\begin{equation}
    S_{int}[\omega,\Psi] = \int \Psi^{A(2s-t)}\,\Psi^{A'(t)}\,
    \hat e_{AA'}\wedge\omega_{A(2s-t-1),A'(t-1)}
\end{equation}
made out of the current associated with the previous pair
of massless fields and the partially-massless field. Note that
we will restrict ourselves to bosonic fields, and hence
will assume that $t$ is even.
As already argued before, all of these pieces are invariant under
shift symmetry. Moreover, the free action is invariant under 
the differential gauge symmetry
\begin{equation}
    \delta_\epsilon\omega_{A(2s-t-2)} = \nabla\epsilon_{A(2s-t-2)}\,,
    \qquad 
    \delta_\epsilon\omega_{A'(t-2)} = \nabla\epsilon_{A'(t-2)}\,,
\end{equation}
for the massless fields, and 
\begin{equation}
    \delta_\xi\omega_{A(2s-t-1),A'(t-1)} = \nabla\xi_{A(2s-t-1),A'(t-1)}\,,
\end{equation}
for the partially-massless field. Under this last gauge transformation,
the variation of the current interaction term reads
\begin{equation}
    \delta_\xi S_{int}[\omega,\Psi] = \int \nabla(\Psi^{A(2s-t)}\,
    \Psi^{A'(t)})\,\hat e_{AA'}\,\xi_{A(2s-t-1),A'(t-1)}\,,
\end{equation}
and vanishes only on-shell as explained before. It can be compensated
off-shell by deforming the gauge symmetry of the pair of massless
fields as follows,
\begin{subequations}\label{currentsymm}
    \begin{align}
        \delta_\xi\omega_{A(2s-t-2)} &= +\tfrac32\,\alpha\,\Psi^{A'(t)}\,e\fud{B}{A'}\,
        \xi_{A(2s-t-2)B,A'(t-1)}\,,\\
        \delta_\xi\omega_{A'(t-2)} & = -\tfrac32\,\alpha\,\Psi^{A(2s-t)}\,e\fdu{A}{B'}\,
        \xi_{A(2s-t-1),A'(t-2)B'}\,,
    \end{align}
\end{subequations}
i.e. with terms depending on the gauge parameter of
the partially-massless field. The variation of the free
actions for the massless fields under this modification
of their gauge symmetry then reads
\begin{equation}
    \begin{aligned}
        \delta_\xi S_{free}[\omega,\Psi]
        & = -\tfrac32\,\alpha\,\int \nabla\Psi^{A(2s-t)}\,
        \Psi^{A'(t)}\,H_{AA}\,e\fud{B}{A'}\,
        \xi_{A(2s-t-2)B,A'(t-1)}\\
        & \quad {+} \tfrac32\,\alpha\,
        \int \Psi^{A(2s-t)}\,\nabla\Psi^{A'(t)}\,H_{A'A'}\,
        e\fdu{A}{B'}\,\xi_{A(2s-t-1),A'(t-2)B'}\,,
    \end{aligned}
\end{equation}
which, upon using
\begin{equation}
    H_{AA}\,e\fud{B}{A'} = +\tfrac23\,\hat e_{AA'}\,\delta^B_A\,,
    \qquad 
    H_{A'A'}\,e\fdu{A}{B'} = -\tfrac23\,
    \hat e_{AA'}\,\delta^{B'}_{A'\,,}
\end{equation}
can be brought to the form
\begin{equation}
    \delta_\xi S_{free}[\omega,\Psi]
    = -{\alpha}\int \nabla\,(\Psi^{A(2s-t)}\,\Psi^{A'(t)})\,
    \hat e_{AA}\,\xi_{A(2s-t-1),A'(t-1)}\,,
\end{equation}
so that the full action \eqref{eq:current_interaction}
is gauge invariant. Note that the deformations \eqref{currentsymm} of the gauge symmetries are Abelian, which is not the case for the current interactions in the non-chiral formulation. A straightforward generalization of these current interactions is to take advantage of other conserved currents that involve derivatives, see e.g. \cite{Gelfond:2006be}. Schematically they read $J_{2s_1+k,2s_2+k}\sim\Psi_{2s_1} \nabla^{k}\bar \Psi_{2s_2}$. In all these cases, except for $s_1=s_2=0$, the action does not require any higher order corrections. 

Note also that this type of interaction is simply a Noether coupling, which is similar
to the one explored in \cite{Boulanger:2019zic}. The spectrum
of the two resulting theories are however different: 
here, we find interactions between a partially-massless
field of spin-$s$ and even depth-$t$, and two massless
fields of spin $s-\tfrac t2$ and $\tfrac t2$, whereas
the interacting theory constructed in \cite{Boulanger:2019zic}
involves only partially-massless spin-$2$ fields. 

\section{Discussion and Conclusions}
\label{sec:disco}
We have studied the simplest types of interactions: Yang--Mills
and current ones. It would be interesting to classify all possible
interactions within the new approach to partially-massless fields
advocated in the present paper. For example, there should exist
partially-massless theories featuring gravitational interactions.
Another important omission is to have genuine non-Abelian higher spin higher derivative interactions. 
Such interactions, as different from, say, the Yang--Mills ones,
introduce nontrivial constraints that fix the spectrum of a theory
together with all the couplings.  

The elephant in the room is twistor theory, which played an important,
but silent, r\^ole in the paper. Indeed, the twistor approach directly
leads to field variables $\Psi^{A(2s)}$ and $\omega^{A(2s-2)}$
for massless fields \cite{Hitchin:1980hp}. This was the starting point
of our generalization to partially-massless fields. However,
the original twistor formulation of partially-massless fields
seems to be missing at the moment. It would be interesting
to bridge this gap. 

At least for the purely massless case there exists a complete,
local higher spin gravity --- Chiral Theory 
\cite{Metsaev:1991mt,Metsaev:1991nb,Ponomarev:2016lrm,Skvortsov:2018jea},
which in addition to Yang--Mills and gravitational interactions
incorporates genuine higher spin interactions. The theory admits
any value of the cosmological constant, including zero. As was shown
in \cite{Ponomarev:2017nrr}, Chiral Theory has two contractions
where the scalar field can be dropped while either Yang--Mills
or gravitational interactions are kept (no genuine higher spin
interactions are present). These two contractions have simple
covariant actions \cite{Krasnov:2021nsq} and twistor origin
\cite{Tran:2021ukl,Herfray:2022prf,Adamo:2022lah}. Within AdS/CFT duality, Chiral Theory
should be dual to a subsector of Chern--Simons matter theories
\cite{Sharapov:2022awp}. 

In view of the facts collected here-above, it looks plausible
that there exist (Chiral) higher spin gravities with partially-massless
fields in the spectrum \cite{Sharapov:2022awp}. These theories should admit
contractions that feature either Yang--Mills or gravitational
interactions, the former of which are considered in the present paper.
Within AdS/CFT duality, such theories should be dual to a subsector
of isotropic (Chern--Simons) Lifshitz CFT's \cite{Bekaert:2013zya},
i.e. of vector models with higher derivative kinetic terms.\footnote{Chern--Simons extension of these models have not been explored so far. It also remains unclear if the $3d$ bosonization duality can be extended to these models.} 

Lastly, it would be interesting to explore a family of deformations
of the actions proposed in the paper via the $\Psi^2$-terms.
Such deformation mimics the well-known result on how Yang--Mills theory
can be represented as a deformation of the self-dual Yang--Mills theory
\cite{Chalmers:1996rq}: $\Psi F(\omega)$-type actions need to be completed
with $\Psi^2$-terms. This idea can be interesting already for free fields,
resulting in a new second order action for partially-massless fields,
which is still simpler than its cousins in terms of non-chiral field
variables. For massless fields the $\Psi^2$-deformation was also shown
to give higher spin theories with nontrivial scattering already in flat space
\cite{Adamo:2022lah}.

\section*{Acknowledgments}
We are grateful to Euihun Joung and Kirill Krasnov  for very fruitful discussions. 
E.S.\ is Research Associate of the Fund for Scientific Research (FNRS), Belgium. The work of S.D. was supported by the European Research Council (ERC) under the European Union’s Horizon 2020 research and innovation programme (grant agreement No 101002551). The work of T.B. was supported by the Fonds de la Recherche Scientifique --- FNRS under Grant No. F.4544.21 and the European Union’s Horizon 2020 research and innovation program under the Marie Sk\l{}odowska Curie grant agreement No 101034383.

\appendix

\section{Beyond maximal depth}
\label{app:beyond}
As is clear from the discussion in Section \ref{sec:free},
the action \eqref{eq:free_PM_action} and equations of motion
\eqref{eq:free_EOM} are formally well-defined beyond the maximal
depth $t=s$. Moreover, the number of physical degrees of freedom
still follows the $2t$-track. While it is beyond the scope of the present
paper to analyze the $t>s$ case in detail, let us make few remarks. 

For $t=s+1$, we are presented with the puzzle
that the $0$-form $\Psi^{A(s),A'(s)}$ is balanced, and hence
in vector language corresponds to a symmetric tensor.
It therefore cannot be related to any Weyl tensor, since
the latter are always valued in two-row diagrams.
For $t=s+2,\dots,2s-1$, let us define $t=2s-\tau$,
with $\tau=1,\dots,s-2$, so that the pairs of fields 
in these cases take the forms
$(\omega^{A(\tau-1),A'(2s-\tau-1)},\Psi^{A(\tau+1),A'(2s-\tau-1)})$.
In this parametrization, the $1$-form $\omega$ seems like
the anti-self-dual part of the last connection for a spin-$s$ field
of depth-$\tau$, but the $0$-form does not have the required
symmetry to be considered as the corresponding Weyl tensor.
This can be traced back to the fact that we used the self-dual
basis $2$-forms $H_{AA}$ in the action to contract the $0$-form
$\Psi$. Consequently, the number of unprimed indices in $\omega$
and $\Psi$ differs by $2$, but when crossing the boundary
$t=s+1$, this difference is now the source of the mismatch 
between the pairs of indices for them to be identified with
the anti-self-dual part of the last connection and Weyl tensor
for a partially-massless field.

More importantly, the equations of motion obtained in these cases
do not describe the propagation of a partially-massless field:
one can check that the first few descendants of the Weyl tensor
which are not constrained by Bianchi identities do not generate
the usual module of a PM anti-self-dual Weyl tensor. Indeed,
consider a $0$-form $\Psi^{A(t-1),A'(2s-t+1)}$ where the parametrization
of its indices suggests that it corresponds to the anti-self-dual
part of the Weyl tensor of a spin-$s$ and depth-$t$ PM field,
subject to the equation of motion
\begin{equation}
    H^{BB}\,\nabla\Psi_{A(t-3)BB,A'(2s-t-1)} \approx 0\,.
\end{equation}
Then, one finds 
\begin{equation}\label{eq:wrong}
    \nabla\Psi_{A(t-1),A'(2s-t+1)}
    = e\fud{B}{A'}\,\Psi_{A(t-1)B,A'(2s-t)}
    + e^{BB'}\,\Psi_{A(t-1)B,A'(2s-t+1)B'}\,,
\end{equation}
instead of
\begin{equation}\label{eq:right}
    \nabla\Psi_{A(t-1),A'(2s-t+1)}
    = e\fdu{A}{B'}\,\Psi_{A(t-2),A'(2s-t+1)B'}
    + e^{BB'}\,\Psi_{A(t-1)B,A'(2s-t+1)B'}\,,
\end{equation}
as would be expected for the anti-self-dual part
of a spin-$s$ and depth-$t$ Weyl tensor. One can notice
that, though the second term on the right hand side of
these two expressions are identical, the first one is not.
In vector language, the expected spectrum of $0$-forms
is given by Young diagrams of the Lorentz group of the form
\begin{equation}
    \Yboxdim{12pt}\scriptstyle
    \gyoung(_9{s}_4{n},_6{s-t+1}_3{m})\,,
\end{equation}
with $n\geq0$ and $m=0,\dots,t-1$. This simply corresponds
to the fact that the derivatives of the Weyl tensor that
are unconstrained by equations of motion and Bianchi identities
are those projected in the first two rows of the Weyl tensor
Young diagram (in arbitrary number in the first row, or only up to $t-1$ in the second row). The equation \eqref{eq:wrong}
is not compatible with this because the two $0$-forms
appearing on the right hand side correspond to the diagrams
\begin{equation}
    \Yboxdim{12pt}\scriptstyle
    \gyoung(_9{s},_5{s-t};{\times})
    \qquad \qquad
    \gyoung(_9{s};,_5{s-t+1})
\end{equation}
so that in particular, the first diagram is unexpected
(see \cite{Skvortsov:2006at,Ponomarev:2010st}), due to the fact
that a box has been removed in the second row (crossed
hereabove) instead of being added. Due to this early
departure in the descendants of $\Psi$, the whole module
generated by the infinite tower of $0$-form required
to build an FDA will not correspond to that of a PM Weyl 
tensor. Once again, this can be traced back to the fact
that the expected equations \eqref{eq:right} is
the parametrization of a generic element in the kernel
of the symplectic form determined by $H_{A'A'}$, i.e.
it is a solution of
$H^{B'B'}\Psi_{A(t-1),A'(2s-t-1)B'B'} \approx 0$.

\begin{figure}[!ht]
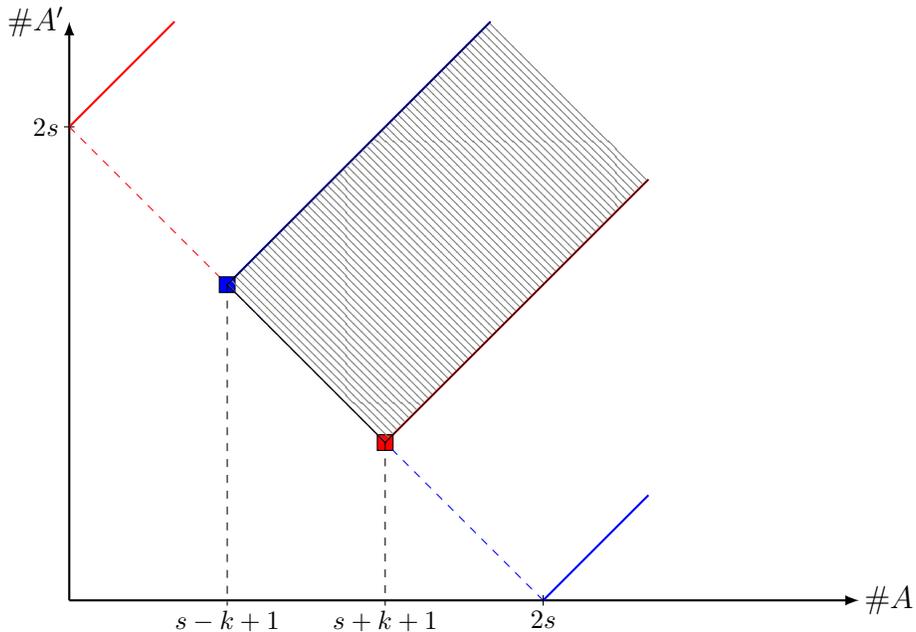

    \FIELDSBeyond
    \label{fig:beyond}
    \caption{In blue, the region covered by descendants
    of $\Psi^{A(s-k+1),A'(s+k-1)}$, in red the descendants
    of $\Psi^{A(s+k-1),A'(s-k+1)}$ and in gray the overlap
    between these two regions.}
\end{figure}

A possible scenario would be that this system, for $t=s+k$
and $k=1,\dots,s-1$, describes a reducible representation
of $\mathfrak{g}_\Lambda$, composed of two massive fields
of spin-$s$ and $k-1$. A trivial, but necessary, check
is that the counting of degrees of freedom is consistent,
since $2t = 2s+1 + 2(k-1)+1$. A more significant hint,
which motivates our conjecture, is that the spectrum
of $0$-forms in this case, represented in Fig. \ref{fig:beyond},
agrees with this proposal. Indeed, when the depth $t$
goes beyond $s$, the two strips of $0$-forms start
overlapping. The whole
region covered by these strips corresponds to the spectrum
of $0$-forms of a massive spin-$s$ field \cite{Ponomarev:2010st},
when each $0$-form appears with multiplicity $1$. 
The overlapping region could similarly be interpreted as
the collection of $0$-forms describing a massive spin-$(k-1)$
field, due to the width of this strip, but that would be represented
by spin-tensors of higher ranks than expected. In other words,
this massive spin-$(k-1)$ field could appear in our system
as a spin-tensor, which, due to some equation of motion,
should be expressed as derivative of a lower rank spin-tensor,
the latter being the genuine massive spin-$(k-1)$ field.
Note that this is to be taken, for the time being, only as
a proposal since proving rigorously the above statement
would go beyond the scope of this paper, and is left for
potential future work.

\footnotesize
\providecommand{\href}[2]{#2}\begingroup\raggedright\endgroup

\end{document}